# 1. Introduction

It is becoming a central strategy in the chemical industry to increase the use of renewable energy as a replacement for fossil fuels, to become more resource and energy efficient. One of the key molecules considered to continue on this road is $CO_2$. The reactions which are of greatest interest are those leading to the production of short-chain olefins (ethylene, propylene) and the conversion of $CO_2$ to syngas, formic acid, methanol and dimethyl ether, hydrocarbons via Fischer-Tropsch Synthesis and methane [1].

At first sight, the reactions in which $CO_2$ is involved can be divided into two categories, chemicals and fuels. The latter is being considered as the most suited target for the conversion of large volumes of $CO_2$ since its market size is 12-14 times larger than the former. One of the most interesting compounds is methanol, which is positioned exactly in the middle, as it is at the same time a raw chemical and a fuel (in combustion engines and fuel cells) [2]. Moreover, it has been extensively discussed that methanol and dimethyl ether (DME) can play a pivotal role in the energy scenario under the "methanol economy" concept [1].

Methanol synthesis from syngas (CO and $H_2$) is a well-known commercial process, and can also be realized starting from $CO_2$ and $H_2$. The traditional approach consists of a two-catalyst system, such as Cu/oxides, to catalyse the reversed water gas shift reaction, followed by a CO reduction to methanol (a typical catalytic system is Cu/ZnO/Al$_2$O$_3$) [1]. This is however considered to be a 3-step renewable energy process, since first an energy vector, $H_2$, has to be produced from electricity, which is then consumed to produce the desired chemical (renewable energy → electricity → hydrogen → chemical/fuel).

The use of plasmas, on the other hand, could provide us with a more efficient 2-step renewable energy process (renewable energy → electricity → chemical/fuel) when starting from $CO_2$ and $H_2O$. Therefore, we carried out experiments in a dielectric barrier discharge (DBD), as well as extensive modelling and a chemical kinetics analysis. This was achieved in order to obtain a better understanding of the mechanisms related to the reactivity of $CO_2/H_2O$ plasmas and of their conversions into value added products.

## 2. Description of the model

### 2.1. 0D Chemical Kinetics Model

The computational model used in this work to describe the plasma chemistry is a zero-dimensional (0D) kinetic model, called Global_kin, developed by Kushner and coworkers [3]. In this work the 0D plasma chemistry module and the Boltzmann equation module are used. The time-evolution of the species densities is calculated, based on production and loss processes, as defined by the chemical reactions. The rate coefficients of the heavy particle reactions depend on the gas temperature and are calculated by Arrhenius equations. The rate coefficients for the electron impact reactions are a function of the electron temperature, and are calculated in the Boltzmann equation module. Finally, the electron temperature is calculated with an energy balance equation.

### 2.2. Plasma Chemistry Included in the Model

The $CO_2$ chemistry used in this study is adopted from the work of Aerts et al. [4] and the hydrocarbon chemistry from the work of Snoeckx et al. [5], while the $H_2O/O_2$ chemistry was taken from the work of Van Gaens and Bogaerts [6] to take into account the corresponding reactions with $CO_2$. The total chemistry set considers 122 different species, which react with each other in 344 electron impact reactions, 930 ion reactions and 537 neutral reactions. Their corresponding rate coefficients and the references where these data were adopted from are listed in [4-6].

## 3. Experimental

The experiments are carried out in a coaxial DBD reactor (see Fig. 1). A stainless steel mesh (ground electrode) is wrapped over the outside of an alumina tube with an outer and inner diameter of 30 and 26 mm, respectively. A copper rod with a diameter of 22 mm is placed in the centre of the alumina tube and used as high voltage electrode. The length of the discharge region is 100 mm with a discharge gap of 2 mm, giving rise to a discharge volume of 15.1 cm3. The DBD is supplied with an AFS generator G10S-V for a maximum power of 1000 W, with peak-to-peak voltage of 5 kV and frequency of 28.06 kHz. The Q-U Lissajous method is used to calculate the discharge power. The energy input is defined as the SEI (specific energy input), which is equal to the ratio of the calculated plasma power to the gas flow rate.

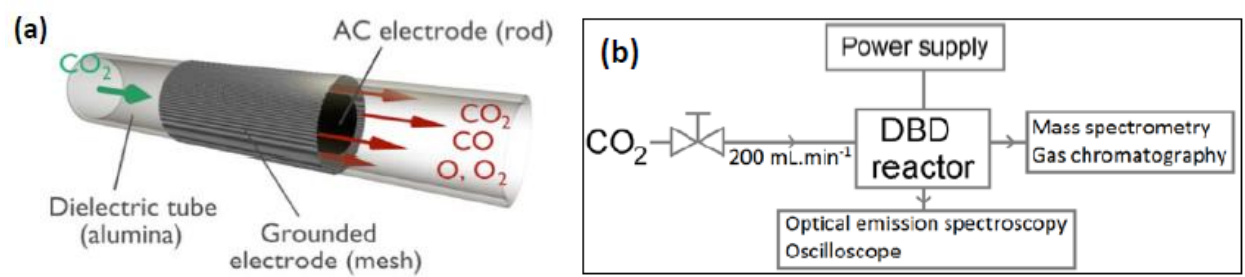

*Fig. 1. Schematic diagrams of (a) the DBD reactor, and (b) experimental setup.*

$CO_2$ is used as feed gas with a flow rate of 250 and 500 $mL_n.min^{-1}$ with a continuous flow of water vapour. This water vapour is generated in a controlled manner using a steam generator (CEM mixer Bronkhorst). Between 0 and 12 % of water vapour was added to the $CO_2$ plasma. Furthermore, the entire system is heated up to 50 °C to avoid condensation and to promote nebulization of the water through the discharge. The $CO_2$ conversion is studied using mass spectrometry (Hiden Analytical QGA MS) and optical emission spectroscopy (Andor Shamrock 500i OES), while electrical characterisation is performed by means of an oscilloscope (Tektronix DPO 3032) to evaluate the properties of the discharge. A small amount of $H_2$ is always observed by mass spectrometry when the plasma contains $H_2O$.

## 4. Results

First, we will discuss the experimental results (section 4.1), explaining the effect of the water vapour and the residence time on the $CO_2$ and $H_2O$ conversion. Subsequently we will compare these experiments with our modelling results, based on reactant conversion and product selectivity. This allows us to use the plasma chemistry in the model to describe and explain the observed trends (section 4.2).

### 4.1. Experimental Results

In Fig. 2 the experimental $CO_2$ and $H_2O$ conversions are plotted as a function of water vapour percentage for a $CO_2$ flow rate of 250 $mL_n$/min. From these results it is clear that the $CO_2$ conversion is always the highest for pure $CO_2$, when no water vapour is added to the discharge. This behaviour may result from the destabilization of the discharge induced by the presence of water, since water has the tendency to trap free electrons. When going from 0 to 4 % water vapour the $CO_2$ conversion drops by a factor 2 for all SEI values investigated. When adding water vapour up to 12 % both the $CO_2$ and $H_2O$ conversion continue to decrease slightly by 20-30 % and 10-20 %.

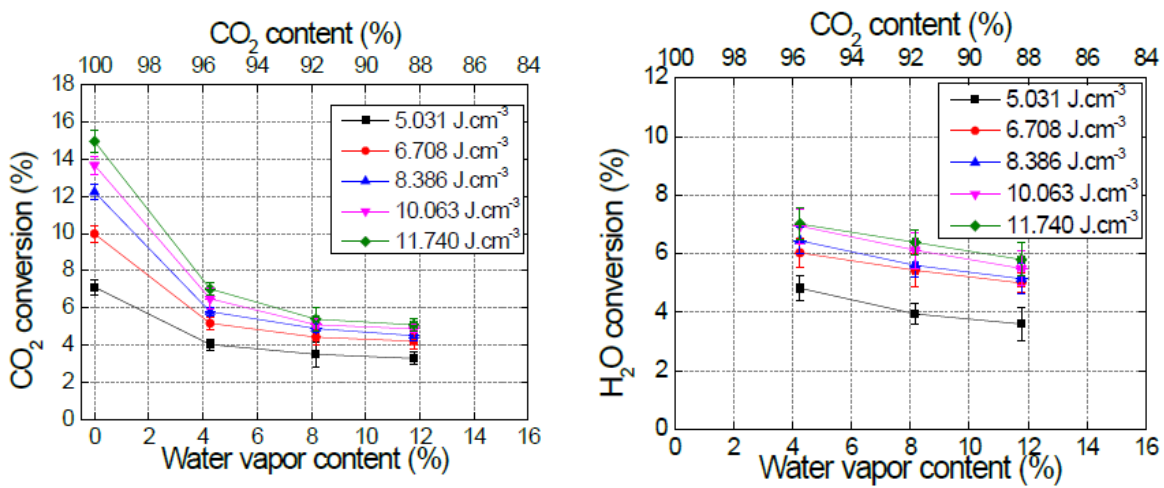

Fig. 2. $CO_2$ (top) and $H_2O$ (bottom) conversion obtained by MS as a function of water vapour content for $CO_2$ flow rate = 250 $mL_n.min^{-1}$.

By increasing the flow rate from 250 $mL_n$/min to 500 $mL_n$/min, the residence time drops by a factor 2, thus the exposure time of the gas molecules to the discharge is shorter, and both the $CO_2$ and $H_2O$ conversion decrease, as can be seen in Fig. 3. The presented results show that the $CO_2$ and $H_2O$ conversion increase when the energy density, i.e. higher SEI, is applied for both $CO_2$ flow rates under study (see Figs. 2 and 3). For all investigated cases the main products formed are CO, $H_2$ and $O_2$.

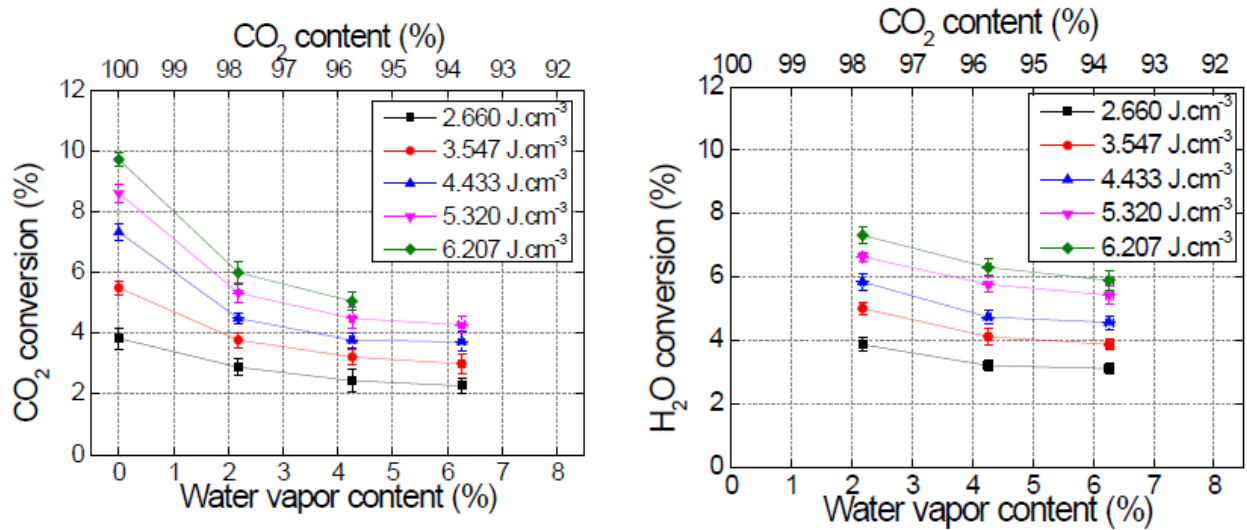

*Fig. 3. $CO_2$ (top) and $H_2O$ (bottom) conversion for obtained by MS as a function of water vapour content for $CO_2$ flow rate = 500 mL$_n$.min$^{-1}$.*

The increase in conversion with increasing SEI is however not strong enough to compensate for the higher energy use, resulting in a decrease of the energy efficiency. This effect is in line with other investigations, albeit for different gas mixtures [7-10]. In summary, the $CO_2$ and $H_2O$ conversion show higher values at low concentrations of $H_2O$ and high SEI, while the energy efficiency is higher at low SEI and low $H_2O$ concentrations.

## 4.2. Analysis of the Plasma Chemistry

To obtain a one-to-one comparison between the experiments and our simulations, we performed simulations mimicking the exact experimental conditions. The same trends were observed as for the experiments with regard to the conversion of $CO_2$, $H_2O$ and the selectivity towards CO, $H_2$ and $O_2$ (currently, the calculations are not all finished yet, but in the presentation, the calculation results will be compared in detail with the experimental data). This allows us to use the plasma chemistry in the model to describe and explain the observed trends. The kinetic analysis reveals that the most important process is the reaction between CO and OH:

CO + OH → H + $CO_2$    $k = 5.4 \times 10^{-14}{}_{[cm3/molecule/s]} (T_{[K]}/298)^{1.50} \cdot \exp(250_{[K]}/T)$

This reaction controls the ratio between the conversion of $CO_2$ and $H_2O$. To explain this in a very simple way, the following will be the main reaction path taking place:

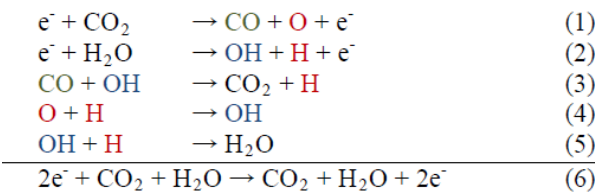

$$
\begin{aligned}
e^- + CO_2 &\rightarrow CO + O + e^- & (1) \\
e^- + H_2O &\rightarrow OH + H + e^- & (2) \\
CO + OH &\rightarrow CO_2 + H & (3) \\
O + H &\rightarrow OH & (4) \\
OH + H &\rightarrow H_2O & (5) \\
\hline
2e^- + CO_2 + H_2O &\rightarrow CO_2 + H_2O + 2e^- & (6)
\end{aligned}
$$

Reactions (1) and (2) lead to the (electron impact) dissociation of $CO_2$ and $H_2O$, yielding the products OH and CO. However, due to the fast reaction rate constant of (3) these radicals will recombine to form again $CO_2$. Thus, two H atoms and one O atom are formed, and as explained before [10], these atoms recombine quickly to form OH and $H_2O$ through reactions (4) and (5), respectively, which are both very fast reactions as well. In the end, this leaves us exactly where we started (6). This is the reason why the conversion of $CO_2$ decreases when $H_2O$ is added and especially why no production of methanol is observed. Indeed, all the hydrogen atoms that are needed to start forming CH and CHO fragments are being steered to OH and subsequently $H_2O$ again.

## 5. Conclusions

We demonstrated that adding water to a $CO_2$ plasma in a DBD leads to a steep decrease in the $CO_2$ conversion, and when adding even more water both the $CO_2$ and $H_2O$ conversion keeps decreasing slightly. As observed for other $CO_2$ mixtures, the conversion increases with increasing SEI, resulting from a decreasing residence time or increasing power. The energy efficiency shows the opposite trend and thus increases with decreasing SEI. The main products formed are CO, $H_2$ and $O_2$, and no methanol formation was observed experimentally. We were able to match the experimental results with our model calculations for an extensive chemistry set. The kinetic analysis of our model revealed why the $CO_2$ conversion decreases when adding water and especially why there was no methanol formation observed. In general, the main reactive species formed in the plasma are OH, CO, O and H. The problem is that the fastest reactions are the recombination reactions of OH and CO to $CO_2$ and H and the recombination reactions of O and H to OH and subsequently $H_2O$.

As we are able to correlate the lower $CO_2$ conversion with these reactions, this allows us to look for possible solutions. When combining the plasma with a catalyst, we should look towards a catalytic system, which is for example able to recombine the present H atoms to molecular hydrogen before it has the chance to recombine to OH and $H_2O$. Also a catalyst which is able to transform the CO together with $H_2$ to methanol before the CO recombines with OH to $CO_2$, would be interesting.

## 6. Acknowledgements

The authors acknowledge financial support from the IAP/7 (Inter-university Attraction Pole) program 'PSI-Physical Chemistry of Plasma-Surface Interactions' by the Belgian Federal Office for Science Policy (BELSPO) and from the Fund for Scientific Research Flanders (FWO). The calculations were performed using the Turing HPC infrastructure at the CalcUA core facility of the Universiteit Antwerpen, a division of the Flemish Supercomputer Center VSC, funded by the Hercules Foundation, the Flemish Government (department EWI) and the Universiteit Antwerpen.